\documentclass[9pt,a4paper,onecolumn]{article}
\usepackage[utf8]{inputenc}
\usepackage{libertine}
\usepackage{amsmath}
\usepackage{amsfonts}
\usepackage{bm}
\usepackage{listings}
\usepackage{graphicx}
\usepackage[pdfa,pdfproducer={},pdfcreator={}]{hyperref}
\usepackage{url}

\usepackage{mathtools}

\usepackage{xcolor}
\usepackage{colortbl} 

\definecolor{bleudefrance}{rgb}{0.0, 0.28, 0.67}
\definecolor{bleudefrance}{rgb}{0., 0.37, 0.67}
 \definecolor{resaltado}{rgb}{0.639, 0.039, 0.961}

\definecolor{histoblue}{HTML}{3981bf}
\definecolor{histobrown}{HTML}{99886e}
\definecolor{histoyellow}{HTML}{c2c099}

\definecolor{miGris}{rgb}{0.5, 0.5, 0.5}

\definecolor{verdevida}{RGB}{108,193,76}
\definecolor{verdevida2}{RGB}{56,100,39}
\hypersetup{
	colorlinks = true,
	citecolor ={bleudefrance},
	allcolors= {bleudefrance},
	linkbordercolor = {white},
}

\usepackage{tikz}

\newsavebox{\circulotachadohorizontalbox}
\sbox{\circulotachadohorizontalbox}{
	\begin{tikzpicture}[baseline=-0.5ex]
		\fill[miGris] (0,0) circle (0.5ex);
		\draw[miGris,line width=0.8pt] (-0.85ex,0) -- (0.85ex,0);
	\end{tikzpicture}
}

\newcommand{\var}{\operatorname{Var}}
\newcommand{\cov}{\operatorname{Cov}}

\usepackage[font=small,justification=RaggedRight]{caption}

\usepackage{pgfgantt}

\usepackage{microtype}
\usepackage[small,compact]{titlesec}

\titleformat{\section}{\bfseries\sffamily\scshape\color{black}}{\arabic{section}}{1em}{\centering\MakeUppercase}

\titleformat{\subsection}{\raggedright\bfseries\sffamily\scshape\small}{\arabic{section}.\arabic{subsection}}{1em}{\MakeUppercase}
\titleformat{\subsubsection}{\centering\bfseries\sffamily\scshape\footnotesize}{\arabic{section}.\arabic{subsection}.\arabic{subsubsection}}{1em}{\MakeUppercase}

\titlespacing{\subsubsection}{0pt}{*4}{*1}
\titlespacing{\subsection}{0pt}{*4}{*1}
\titlespacing{\section}{0pt}{*5}{*2}

\usepackage{fancyhdr}
\usepackage[left=1.50cm, right=1.50cm, top=1.5cm, bottom=2.00cm]{geometry}
\usepackage{amsmath}

\usepackage[backend=bibtex,isbn=true,url=true,eprint=false, bibstyle=numeric,sorting=none,    style=numeric,punctfont=true,maxbibnames=10, minbibnames=10]{biblatex}

\AtEveryBibitem{%
	\clearfield{note}%
}

\DeclareCiteCommand{\citenum}
{}
{\printfield{labelnumber}}
{}
{}

\bibliography{bibs}	

\usepackage{ragged2e}
\usepackage{booktabs}
\usepackage{tabularx}

\usepackage{placeins}

\usepackage[misc]{ifsym}

\usepackage{float}
\restylefloat{table}

\usepackage{appendix} 

\usepackage{abstract}
\usepackage{orcidlink}

\begin{document}

	\title{\Large\textsf{\textbf{{{Evaluating the Uncertainty in Mean Residual Times: Estimators Based on Residence Times from Discrete Time Processes}}}}}

	\author{\small \textsf{\scshape{Hernán R. Sánchez}}$^{1}${\scriptsize\Letter}  \orcidlink{0000-0003-1058-3396}, \textsf{\scshape{Javier Garcia}}$^2$            \orcidlink{0000-0002-3264-4359}\\   }	
	\date{}
	
	\maketitle
	
	\begin{center}
	{\vspace{-0.5cm}\small 
		 $^1$Instituto de Física de Líquidos y Sistemas Biológicos,  UNLP-CONICET, La Plata, 1900, Argentina

				 $^2$Instituto de F\'isica La Plata, UNLP-CONICET, C.C.67, La Plata 1900, Argentina
		
		\footnotesize {\scriptsize\Letter} \textsf{\href{mailto:hernan.sanchez@quimica.unlp.edu.ar}{hernan.sanchez@quimica.unlp.edu.ar}}
		}
\end{center}

	\begin{onecolabstract}
 \vspace{-0.3cm}
 
In this work, we propose estimators for the uncertainty in mean residual times that require, for their evaluation, statistically independent individual residence times obtained from a discrete time process.  We examine their performance  through numerical experiments involving well-known probability distributions, and an application example using molecular dynamics simulation results, from an aqueous NaCl solution, is provided. These computationally inexpensive estimators, capable of achieving very accurate outcomes, serve as useful tools for assessing and reporting uncertainties in mean residual times  across a wide range of simulations.
  		\vspace*{0.8cm}
	\end{onecolabstract}

\section{Introduction}

Across a wide array of knowledge domains, it is crucial to characterize the time that entities of a particular type dwell in a specific spatial region. When the duration of these residencies is random, the mean residence time (mRT) and the mean residual time (mrT) are the metrics most commonly reported to this end. This is frequently seen in studies involving reactors in engineering\cite{scott2016elements,Nauman2008}, water dynamics in biophysics\cite{rahaman2015,Qaisrani2019,Sheu2019,Kemmler2019}, and drug discovery in pharmacology\cite{copeland2006drug,bernetti-farmacos,moschen2016,Weiss2005TheRO}. Analogous concepts are also found in other domains such as medicine\cite{jin2020generalized,Wenwen2023}, oceanography\cite{zhang2010simulation,shen2011modeling} and computer sciences\cite{Bitar}. Given the terminological ambiguity in the related literature, we will clarify in simple terms what we refer to as mRT and mrT in this study. We refer to mRT as the average length of stays, and mrT as the average waiting time for the completion of the stays. In practice, the exact values of mRT and mrT are unknown, hence it is desirable to accompany the estimated values with an assessment of the associated uncertainty. This paper focuses on the estimation of the uncertainty associated with the mrT.  Although this study was motivated by the analysis of molecular dynamics simulations, its results are broadly applicable.

In the field of molecular dynamics, the most commonly used techniques for the calculation of mRTs or mrTs do not involve the direct averaging of the RTs nor rTs associated with the studied process. Instead, they typically rely on the calculation of survival or correlation functions from an auxiliary list ($\mathbf X$), indexed by time, of zeros and ones that represent whether the particle of interest is outside or inside the region in question\cite{impey,garcia1993,brunne1993,laage2008}. The functions are then fitted to a functional form, typically exponential, and the mRT and mrT are derived from the fitting parameters.

Obtaining a list of successive RTs from $\mathbf X$ is not computationally expensive. Usually, residence times behave as identically distributed random variables that exhibit very low correlation among themselves. This allows to estimate the variance of the mRT by dividing the variance of the RTs by the number of variables ($N$).
For a sample collected at regular time intervals, each RT value has a fixed number of corresponding rT values.
From a list of RTs, it is straightforward to obtain a list of rTs. Unlike RTs, rTs are usually correlated, so the dispersion of their mean cannot be calculated in the same manner as that of the mRT. The mrT can be expressed in terms of the RTs and can be efficiently calculated as done in reference \citenum{Sanchez2022}, \S 2.4.   

In the present work, by assuming that the RTs are independent and identically distributed variables (i.i.d) we derive from such expression estimators for the variance of the mrT. We evaluate their performance by testing them against random samples of RTs generated from different probability distributions.  As an illustration of their broad utility, we employed our estimators on molecular dynamics simulation data to estimate the variance of the mrT for water molecules within the first solvation shell of aqueous chloride ions.

\section{Methods}
We will perform the development in terms of number of time steps, rather than times. Specifically, we will assume that the samples are taken at equally spaced discrete times, and the time step is denoted by $\Delta t$. In this case each RT can be written as $t_i = x_i \Delta t$, where $x_i$ is a natural number representing the number of steps of the $i$-th residence time. This approach avoids the need for multiplications by $\Delta t$, leading to slightly simpler expressions.
Considering this and following reference \citenum{Sanchez2022}, \S 2.4, we can express the mean of the number of residual steps as the following statistic:
\begin{equation}
	f_N({\mathbf x}) = \frac{1}{2}+\frac{1}{2} \frac{\sum_{i=1}^{N} x_i^2}{\sum_{i=1}^{N}x_i},
	\label{eq:mrt}
\end{equation}

\noindent where $\mathbf x:=[x_1,x_2,\dots,x_N]$ being $N$ the number of RTs included.  It is worth noting that if times were used instead of time steps, the resulting estimator would be the same, except multiplied by $(\Delta t)^2$. We employed two methods to propose estimators based on the expression presented in Eq. \ref{eq:mrt}, assuming the $x_i$'s are i.i.d.. No assumptions on the discrete probability distribution followed by the RTs was made in the derivations, which implies that the proposed estimators possess a high degree of generality. Detailed descriptions of these methods can be found in the subsequent subsections. 
\subsection{First method\label{sec:method1}}
To determine the variance of $f_N$, we begin by expanding it in a Taylor polynomial around a vector $\boldsymbol{\mu}$, where each of the scalar components is equal to the expected value, $\mu$, of the $x_i$'s variables, which share it since they are i.i.d.:
\begin{equation}
	f_N({\mathbf x}) = f_N(\boldsymbol{\mu}) + \sum_{k=1}^{\infty} \left [ \sum_{i_1\ldots i_k}^{N}  \frac{f_N^{(i_1\dots i_k)} (\boldsymbol{\mu}) }{k!}\prod_\alpha^k (x_{i_\alpha} - \mu) \right],
	\label{eq:taylor}
\end{equation}
where
\begin{equation*}
	f_N^{(i_1\ldots i_k)}(\boldsymbol{\mu}) := \left. \frac{\partial^k f_N({\mathbf x})}{\partial x_{i_1}\ldots \partial x_{i_k}}\right|_{{\mathbf x} = \boldsymbol{\mu}}  = \frac{(-1)^k}{N^k \mu^{k-1}} \left[(k-2)! \sum_{\alpha<\beta}^{k} \delta_{i_\alpha i_\beta} - \frac{k!}{2}\right],
\end{equation*}
which can be proved by induction. The coefficients $f_N^{(i_1 \ldots i_k)}(\boldsymbol{\mu})$ depend on the number of pairs of indices $i_\alpha$ that are equal between themselves, but not on the values of the repeated indices; for example, $f_N^{(12)}(\boldsymbol{\mu}) = f_N^{(35)}(\boldsymbol{\mu}) \neq f_N^{(22)}(\boldsymbol{\mu}) = f_N^{(44)}(\boldsymbol{\mu})$.
We apply the variance operator to the expansion \eqref{eq:taylor}, and use the well-known result for the variance of a linear combination, i.e.,
\begin{equation}
	\operatorname{Var} \left(\sum_{i=1}^{N} a_i X_i \right) = \sum_{i=1}^N a_i^2\operatorname{Var}(X_i)+\sum_{i\not=j}a_ia_j\operatorname{Cov}(X_i,X_j),
	\label{eq:variance_property}
\end{equation}
where ${\operatorname{Cov}}$ denotes the covariance operator, defined as ${\rm Cov}(X, Y) = \operatorname{E}[XY] - \operatorname{E}[X]\operatorname{E}[Y]$. In this context, E is the expected value operator, and $X$ and $Y$ are two random variables. 
Then, the variance of $f_N$  satisfies
\begin{equation}
	\var (f_N) = \sum_{kl}^{\infty} \sum_{\substack{i_1\dots i_k \\ j_1\dots j_l}}^{N}  \ \frac{f_N^{(i_1 \ldots i_k)}(\boldsymbol{\mu})}{k!} \frac{f_N^{(j_1\dots j_l)}(\boldsymbol{\mu})}{l!} \, \sigma_{i_1\dots i_k, j_1 \dots j_l},
	\label{eq:variance_taylor}
\end{equation}
where 
\begin{equation}
	\sigma_{i_1\dots i_k, j_1 \dots j_l}
	 := \cov \left(\prod_\alpha^k (x_{i_\alpha} - \mu), \prod_\beta^l (x_{j_\beta} - \mu) \right).
	 \label{eq: definicion sigma}
\end{equation}

Since the $x_i$'s are uncorrelated with each other, we can express $\sigma_{i_1\dots i_k, j_1 \dots j_l}$ in terms of the central moments of their distribution as follows: First we count the number of repetitions of each index. Then, we construct a product with as many factors as there are unique indices. Each factor corresponds to a different index and is equal to the $m$-th central moment ($\mu_m$), where $m$ is the number of repetitions of that index. For example, in $\sigma_{iijj,jjj}$, $j$ appears 5 times and $i$ appears 2 times, yielding  $\mu_5\mu_2$. Next, we repeat this process for the first set of indices $\left\{i_1\dots i_k\right\}$,  and also for the second set $\left\{j_1\dots j_l\right\}$. We then multiply the corresponding products. Finally, we subtract the result from this step from the result obtained in the initial step. For the previous example, we get $\mu_5\mu_2 - (\mu_2^2)(\mu_3)$.

 Note that the lists of indices, $i_1\dots i_k$ and $j_1 \dots j_l$, not only determine the expressions of the covariances but also the corresponding complete terms of Eq. \ref{eq:variance_taylor}.  We significantly reduced the computational cost by exploiting three conditions. The first two, which eliminate terms  from Eq. \ref{eq:variance_taylor}, are as follows:  1)~$\sigma_{i_1\dots i_k, j_1 \dots j_l} = 0$ if none of the indices of the set $\left\{i_1\dots i_k \right\}$ match any of the indices from the set  $\left\{j_1\dots j_l\right\}$, because it implies independence between the first and second arguments of the RHS of Eq. \ref{eq: definicion sigma}.
2) $\sigma_{i_1\dots i_k, j_1 \dots j_l}=0$ if at least one index of the set $\left\{i_1, \dots, i_k, j_1 \dots, j_l\right\}$ is unique. For example: $\sigma_{1114,111} = 0$, and $\sigma_{123,124} = 0$. 
 The third condition allows to reduce the computational cost by grouping terms that are equivalent in the sense that 
 there are no differences between them across all possible cases. 
 Thus, we only need to calculate the value of one term per group. It can be stated as: 3) The order and values of the indices are irrelevant, only the number of times they appear in each subset matters. For example: $\sigma_{1122,222} = \sigma_{4141,111}$.

We automated most of the procedure using Python3\cite{van2014python} scripts to reduce the likelihood of mistakes and to access higher orders than feasible by hand. We thus obtained approximate expressions for the variance using Taylor expansions up to order 8. The estimators proposed in this section are obtained by substituting the central moments of the distribution in these expressions with their corresponding estimators, which are computed from the sample RTs. Due to their extensions and to keep information tractable, these expressions were placed in the Supplementary Information, and were labeled SM being M the highest order included in the Taylor polynomials.

\subsection{Second method}
We also followed another similar approach with which we obtained an estimator that provides good results and has a fairly compact expression in terms of the raw moments. Notice that Eq. \eqref{eq:mrt} can be written as $f_N({\bf x})=1/2+R({\bf x})/(2S({\bf x}))$, where 
\begin{equation*}
R({\bf x}) = \sum_{i=1}^{N} x_i^2
\end{equation*}
and
\begin{equation*}
S({\bf x}) = \sum_{i=1}^{N} x_i
\end{equation*}
Since $R$ and $S$ can be considered random variables, the variance of their ratio can be approximated by the formula\cite{benaroya2005probability}
\begin{equation}
	{\rm Var}\left(\frac{R}{S}\right) \approx \frac{\mu_R^2}{\mu_S^2} \left( \frac{\sigma_R^2}{\mu_R^2} - \frac{2\,{\rm Cov}(R,S)}{\mu_R\mu_S} + \frac{\sigma_S^2}{\mu_S^2}\right),
	\label{eq:var_ratio}
\end{equation}
where $\mu_R$ and $\mu_S$ are the expected values of $R$ and $S$, $\sigma^2_R$ and $\sigma_S^2$ are the variances of $R$ and $S$, respectively. %
Let $\mu_n^\prime$ be the $n$-th raw momenta of the distribution of the $x_i$'s, the quantities referred to in Eq. \eqref{eq:var_ratio} are found to be 
\begin{equation}
	\begin{aligned}
		&\mu_S = N \mu, \quad \sigma_S^2 = N ( \mu_2^\prime - \mu^2 ), \\
		&\mu_R = N \mu_2^{\prime\,2}, \quad \sigma_R^2 = N ( \mu_4^\prime - \mu_2^{`\prime\,2}), \\
		&{\rm cov}(R,S) = N(\mu_3^\prime - \mu_2^\prime\mu).
	\end{aligned}
	\label{eq:values}
\end{equation}
Substituting the expressions from Eqs. \eqref{eq:values} into Eq. \eqref{eq:var_ratio} and dividing by 4, in accordance with Eq. \eqref{eq:variance_property},  yields an approximation of the variance of $f_N$:
\begin{equation}
	{\rm Var}(f_N) \approx \frac{1}{4Nm_1 ^{\prime2}}\left( m_4^\prime - \frac{2 m_2^\prime m_3^\prime}{m_1^\prime} +\frac{m_2^{\prime3}}{m_1^{\prime2}}\right),
	\label{eq:var_estimator_quotient}
\end{equation}
\noindent where 
\[  m_n^\prime=\frac{1}{N} \sum_{i=1} ^N  x_i^n,\]
represents an unbiased estimator of the $n$-th raw moment.

\section{Results and Discussion}
First, we will assess the performance of the proposed estimators by applying them to $x_i$'s sampled from two well-known probability distributions. One of the distributions we chose is  the shifted geometric distribution, as it is well known that commonly the RTs follow a distribution that is close to the exponential one. The other distribution employed is the discrete uniform distribution, which features a constant probability mass function, which is markedly different from that of the geometric distribution. Thus, we acquire insight into the performance of the proposed estimators in very different scenarios.
We present results for these distributions across various parameter sets. 
 
For the shifted geometric distribution, which is defined by the probability mass function (PMF) $\mathbb{P}(x;p)=(1-p)^{x-1}p$, $x\in[1,\infty)$, we considered six distinct values of the parameter $p$: 0.5, 0.1, 0.05, 0.01, 0.005, and 0.001. Similarly, for the discrete uniform distribution, which is defined by the PMF: $\mathbb{P}(x;a,b)=(b-a)^{-1}$, $x\in[a,b]$, we examined six different pairs of parameters $(a,b)$: (1,100), (10,100), (90,100), (1,1000), (100,1000), and (900,1000). 
For each chosen value of $p$ and each chosen pair of parameters $(a,b)$, we generated 1 million sets of pseudo-random samples, using  in each case the following sample sizes $N$: 30, 69, 158, 362, 829, 1902, 4361, and 10000.

Here, the variance will be presented in terms of steps. Values expressed in time units are obtained by multiplying them by $(\Delta t)^2$. This is not relevant for the evaluation of estimators and allows us to avoid inconvenient distinctions between steps and times that may complicate the reading of the manuscript. It should be noted that, throughout the text, we will employ the usual time-related terminology, as the quantities are proportional and their distinction is irrelevant for the purposes of our analysis.

For each combination of parameters and $N$, we computed the mrT using Eq. \ref{eq:mrt} on each of the respective one million sets, and the variance of the results was computed.  These correspond to the reference values we aim to estimate and are plotted using gray lines in the figures. Figure 1 is a graphical representation of the results for the geometric distribution, while Figure 2 is a graphical representation of the results for the discrete uniform distribution. 

In the figures, we include the results of two of the estimators. One from Eq. \ref{eq:var_estimator_quotient}, and the other that uses Taylor polynomials up to order 8, included in the Supplementary Information, Eq. S8. We opted not to include results from lower order expansions in order to simplify the plot, as increasing the expansion to order 8 does not significantly impact computational time. We applied each of these two estimators to each set of one million samples to obtain their respective values. Their means are also represented in the figures with black and red circumferences ({\large$\circ$}), respectively. 

Additionally, we include the results derived from employing the exact moments of the chosen distributions, which can be computed from their moment generating functions. They are represented using the same color association mentioned in the previous paragraph.

It is worth noting that the plotted values have a precision of between 3 and 5 significant figures, approximately, except for those computed using the exact moments, which are not subject to uncertainties. Therefore, the sample size is sufficiently large for the graphical representation to be accurate.

\begin{figure}
	\centering
	\includegraphics[scale=1.]{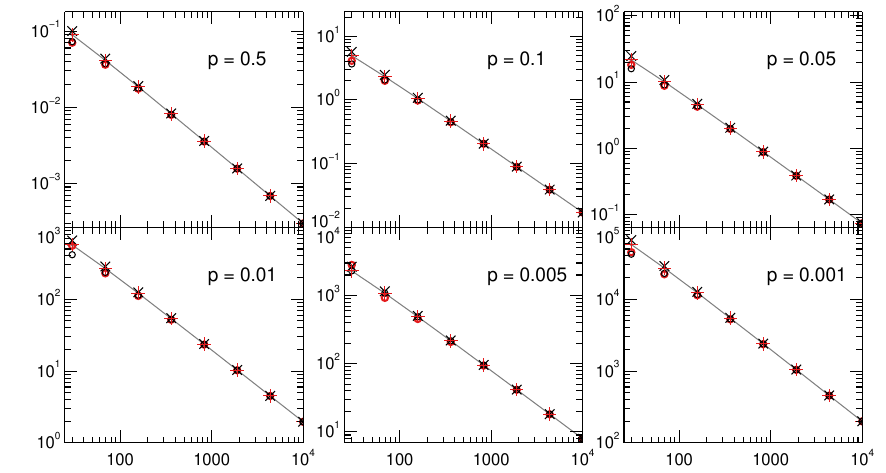}

	\caption{Variance vs. $N$ for the (shited) geometric distribution. \usebox{\circulotachadohorizontalbox}: simulated reference values. {\large$\circ$}: mean of the values obtained from  Eq. \ref{eq:var_estimator_quotient}. {\color{red}\large$\circ$}: mean of the values obtained from  Eq. S8. ${\color{black}\times}$: values from Eq. \ref{eq:var_estimator_quotient} using exact moments.  ${\Large\color{red}+}$: values from Eq. S8 using exact moments.}
	\label{fig:geometric}
\end{figure}

\begin{figure}
	\centering
	\includegraphics[scale=1.]{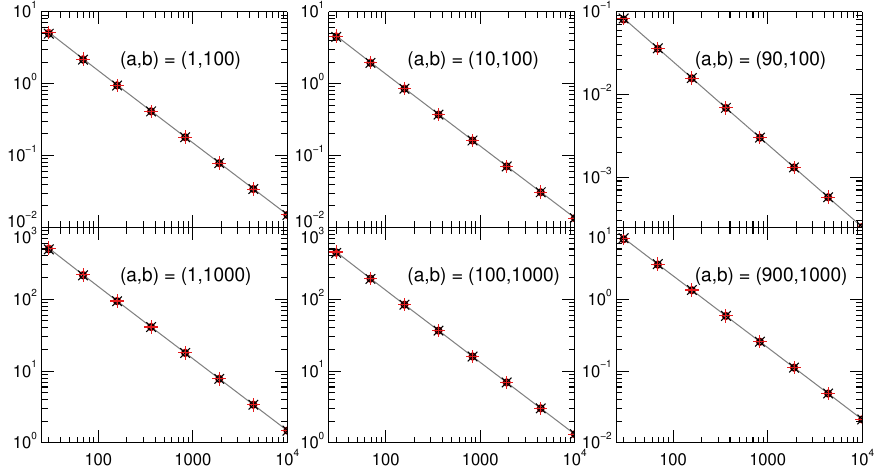}
	\caption{Variance vs. $N$ for the discrete uniform distribution.\usebox{\circulotachadohorizontalbox}: simulated reference values. {\large$\circ$}: mean of the values obtained from  Eq. \ref{eq:var_estimator_quotient}. {\color{red}\large$\circ$}: mean of the values obtained from  Eq. S8. ${\color{black}\times}$: values from Eq. \ref{eq:var_estimator_quotient} using exact moments.  ${\Large\color{red}+}$: values from Eq. S8 using exact moments.}
	\label{fig:uniform}
\end{figure}

The results obtained through Eq. \ref{eq:var_estimator_quotient} are very satisfactory. 
It should be noted that the uncertainty in a measurement is often reported as the standard deviation, not variance. 
Therefore the relative error corresponds to the square root of that implied in the figures. 

For very small values of $N$, the differences between the curves become noticeable, especially in the case of the geometric distribution. For even smaller values of $N$, such as fewer than ten, we observed that truncating Eq. \ref{eq:variance_taylor} at $k=l=8$ is insufficient to obtain converged results. However, this is not a cause for concern, as highly dispersed estimates are inherently expected for very small sample sizes. Therefore, a rigorous study would involve a significantly larger sample size, e.g. hundreds of thousands in the practical case presented in the next section.

It is interesting to note, although it is not of practical use, that for $N=1$ the estimators derived using the first described method are reduced to the expression resulting from applying the variance operator to Eq. $\ref{eq:mrt}$, which can be trivially derived for this particular case. 
As $N$ increases, the sample moments tend to those of the distribution, so the variances obtained using the sample moments tend towards those calculated with the distribution's moments, resulting in less dispersion. The accuracy of both methods increases with $N$. 

For the case of the discrete uniform distribution, the variance can be calculated exactly when $N$ and the interval $(a,b)$ are small.
This can be done directly from the definition of variance and the PMF of the distribution. We computed the variance for various cases and compared these values to the results obtained using the expressions of the estimators, which use the exact moments of the distribution in place of the estimators of the moments. We observed systematic improvements in the results as the order of the polynomials increased, with very high accuracy for higher orders. An example of this is provided in Table \ref{Table}, for the case $N=10$ and $(a,b)=(93,100)$, we obtained correctly the first 14 digits using order 8 and 5 digits for order 3,  It is noteworthy that in internal tests, we found that the results of this last expression are usually of a similar quality to those from the other development up to 3rd order. 

The table also contains the results for the same distribution and parameters when $N=1000$, as well as the results for the geometric distribution with $p=0.05$, for $N=30$ and $N=1000$. Although the computational cost precludes obtaining exact values for these cases, the estimators from Sec. \ref{sec:method1} appear to converge to specific values. The convergence rate accelerates as $N$ increases, and this effect is more pronounced for the geometric distribution. The estimator from Eq. \ref{eq:var_estimator_quotient} yields reasonably accurate approximations, although it is expected that estimators derived from higher-order Taylor polynomials will provide superior accuracy.

\begin{table}[bht!]\small\centering
	\begin{tabular}{lrrrr}
		\cmidrule[1.5pt](lr{0.0em}){1-5}%
		& \multicolumn{2}{c}{{\sf\bf Geometric}} & \multicolumn{2}{c}{{\sf\bf Discrete uniform}} \\
		\cmidrule[0.8pt](lr{0.9em}){1-5}%
		& \multicolumn{1}{c}{\sf30} & \multicolumn{1}{c}{\sf1000} & \multicolumn{1}{c}{\sf10} & \multicolumn{1}{c}{\sf1000} \\
		\cmidrule[0.8pt](lr{0.6em}){2-3}%
		%\cmidrule[0.8pt](lr{0.6em}){3-3}%
		\cmidrule[0.8pt](lr{0.6em}){4-5}%
		%\cmidrule[0.8pt](lr{0.6em}){5-5}%
		\sf\bf	(\ref{eq:var_estimator_quotient}) & 24.70 & 0.74100000 & 0.1311584285189072 & 0.0013115842851890724 \\
		\sf\bf	S1       & 3.17  & 0.09500000 & 0.1312500000000000 & 0.0013125000000000000 \\
		\sf\bf	S2       & 37.80 & 1.18610357 & 0.1313089848049612 & 0.0013130641249664420 \\
		\sf\bf	S3       & 19.25 & 0.73544207 & 0.1311923130039270 & 0.0013115879454227196 \\
		\sf\bf	S4       & 23.12 & 0.73878468 & 0.1311923124294356 & 0.0013115879450302053 \\
		\sf\bf	S5       & 20.96 & 0.73772937 & 0.1311922958779697 & 0.0013115879425383327 \\
		\sf\bf	S6       & 21.84 & 0.73774821 & 0.1311922958770776 & 0.0013115879425383307 \\
		\sf\bf	S7       & 21.31 & 0.73774308 & 0.1311922958733286 & 0.0013115879425383236 \\
		\sf\bf	S8       & 21.61 & 0.73774323 & 0.1311922958733283 & 0.0013115879425383238 \\
		\sc\sf\bf	Exact & \multicolumn{1}{c}{-} &   \multicolumn{1}{c}{-}           & 0.1311922958733272 & \multicolumn{1}{c}{-} \\
		\cmidrule[1.2pt](lr{0.1em}){1-5}%
	\end{tabular}
	
	\caption{Computed values employing the exact moments for geometric ($p=0.05$) and discrete uniform ($(a,b)=(93,1000)$) distributions using various equations and $N$ values ($N=30$, $1000$ for geometric; $N=10$, $1000$ for discrete uniform). The first column refers to the equation used, with ``Exact" denoting values computed without approximations.\label{Table}}
\end{table}

Thus far, we have demonstrated that the estimators presented in this paper yield highly accurate results for uncorrelated samples, provided adequate sample sizes of a few tens or more are available. It performs well for both normal and geometric distributions, the latter of which represents typical situations in simulations.

\subsection{Example of application to a molecular dynamics simulation}\label{sec: practical case}
As an illustration, in this section we employ the estimator from Eq. \ref{eq:var_estimator_quotient} and Eq. S8 in an applied context. A molecular dynamics simulation involving a chloride anion, a sodium cation, and 1000 water molecules is employed, and the study will focus on the residence of oxygen atoms in the inner region of a sphere, of a fixed radius and centered on the chloride.

The simulation ran for 100 ns using a 2 fs time step, and values were recorded at 0.1 ps intervals. Water and ions were modeled using the SPC/E\cite{berendsenspce1987} and OPLS\cite{OPLS} force fields, respectively.   
The system was thermostatted at 298.15 K with the Bussi-Donadio-Parrinello velocity rescaling algorithm\cite{giovanni2007},   and maintained at 1 bar using the Berendsen's barostat\cite{berendsen1984}.
  The system was equilibrated before the production run, and Gromacs 2022.3\cite{abraham2015} was used for the simulation.  
  The radius used is 0.3875 nm, which corresponds to the first local minimum of the radial distribution function. 
  This value was obtained using the MDTraj library\cite{mcgibbon} in custom Python3 scripts.

As mentioned before, the most prevalent methods for computing mRT and related quantities usually involve, for each particle, a binary vector $\mathbf X$ arranged in temporal order, where 1 denotes that the particle is within the region of interest and 0 indicates otherwise. Notably, these methods typically do not require knowledge of each individual residence time, which are needed for the application of the estimator proposed by us. In order to obtain the list of residence times, we apply the algorithm proposed by one of us in reference \citenum{Sanchez2022} (\S 2.4) which indeed computes the residence times. In strict terms, we make a trivial modification to the algorithm, such that the residence times are stored in an array.

The approaches to compute mRTs often include some strategy to avoid considering a residence as finished in the event of brief transient exits from the region of interest. For this purpose, we have used an algorithm also proposed in reference \citenum{Sanchez2022} to preprocess $\mathbf X$.  If the particle is allowed to be absent continuously for up to $k-1$ steps, then the convolution of $\mathbf X$ with a vector of ones of $k$ elements ($\mathbf v$) is computed.  Then the elements of the resultant vector greater than 1 are replaced by 1. The convolution of the newly obtained is taken with $v$. Finally, the elements of the resulting vector are replaced with 0 if they are less than $k$ and with 1 otherwise. As in reference \citenum{Sanchez2022},  the primary objective of employing this method here is to facilitate comparisons. We opt to consider that a particle's residence is terminated when it reaches $t^*=2$ ps continuously outside the region of interest. This value is typical for aqueous solutions\cite{laage2008}. 

By applying the procedures described above, we found over 350,000 RTs in the simulation. This number was reduced to approximately 137,000 RTs after applying the filter for transient escapes. Using Eqs. \ref{eq:mrt} in conjunction with either Eq. \ref{eq:var_estimator_quotient} or Eq. S8, we calculated mrT = 11.06 $\pm$ 0.06 ps where the uncertainty was reported as the standard deviation, that is, the square root of the value obtained using Eq. \ref{eq:var_estimator_quotient} or Eq. S8. It is worth noting that in this case, the relative error of using Eq. \ref{eq:var_estimator_quotient} instead of Eq. S8 for calculating the standard deviation is less than $3\cdot10^{-5}$.
The mRT obtained is 5.899 $\pm$0.026 ps. For computing uncertainties we assumed that the RTs are uncorrelated. To corroborate this assumption,  we computed the autocorrelation function of 1000 vectors containing the temporally ordered residence times of each of the 1000 oxygen atoms. We calculated the mean of their firsts lags and their variance. The second average lag was -0.00829, similar to subsequent ones, and of the same order as their standard deviation, thereby confirming our assumption.

The results are in good agreement with those of reference \citenum{laage2008}. In that methodological work, a similar system was studied, and by applying the method from reference \citenum{impey} with $t^*=2$~ps, the authors obtained a value of 11.6 ps. As demonstrated in reference \citenum{Sanchez2022}, under certain conditions, that method yields what we call mrT here, so our results are similar. They found a value of 7.3 ps when applying their own method to the same system. Such method automatically discards brief transient escapes. The authors argued that the discrepancy in their results was due to an inadequate value of $t^*$ for that system. Our results suggest that, regardless of the suitability of 2 ps as the value of $t^*$ for this system, the reason for the discrepancy could lie in which property is being calculated. It is significant to clarify that our results are internally consistent. One way to corroborate this is by employing the expression of the inspection paradox for a discrete-time process derived in \citenum{Sanchez2022}.
\begin{equation*}
	\rm{mrT} = \frac{\rm{mRT}^2+\rm{Var(t_i)}  }{2\rm{mRT} } + \frac{\Delta t}{2}
\end{equation*}
Employing all the decimals acquired,  we verify the equality up to 15 decimal places. Naturally, this does not encompass the variance of the mrT.

\section{Summary and conclusions}
In this work, we presented estimators for the variance of the mean residual time, Eqs. \ref{eq:var_estimator_quotient} and Eqs. S1-S8. Their estimates rely solely on the RTs samples. These  estimators do not depend on the specific probability distribution followed by the RTs. 
We tested the proposed estimators by sampling from the shifted geometric distribution, which is applicable to many real-world situations, and from the discrete uniform distribution, which is markedly different from the first one. With sufficiently large sample sizes, excellent results were obtained in both cases, demonstrating robustness and accuracy. A practical application was presented in Section \ref{sec: practical case}, where we analyzed the residual times of water molecules around chloride ions in diluted aqueous solution. The successful performance of the estimators highlights their broad utility across distributions and sample sizes.

We conclude that the proposed estimators are useful for computing uncertainties in mean residual times of discrete-time processes from i.i.d. RTs. 
Their implementation is straightforward, and the associated computational cost is negligible.

\section*{Acknowledgement}
This work was supported by the National Scientific and Technical Research Council of Argentina (CONICET).
\linespread{1.1} 
\printbibliography
\end{document}